\documentclass{cccg21}
\usepackage{enumitem,graphicx}

\title{Succinct Euler-Tour Trees}

\author{Travis Gagie%
        \thanks{%
            Faculty of Computer Science,
            Dalhousie University,
            {\tt travis.gagie\,@\,dal.ca}%
        }
	\and
		Sebastian Wild%
		\thanks{%
			Department of Computer Science, 
			University of Liverpool, 
			\texttt{sebastian.wild\,@\,liv.ac.uk}%
		}
}

\index{Gagie, Travis}
\index{Wild, Sebastian}

\begin{document}
\thispagestyle{empty}
\maketitle

\begin{abstract}
We show how a collection of Euler-tour trees for a forest on $n$ vertices can be stored in $2 n + o (n)$ bits such that simple queries take constant time, more complex queries take logarithmic time and updates take polylogarithmic amortized time.
\end{abstract}

\section{Introduction}
\label{sec:introduction}

Tarjan and Vishkin~\cite{TV85} showed how we can efficiently solve several problems in algorithmic graph theory by representing planar embeddings of trees as Euler tours of corresponding directed graphs.  To obtain the graph corresponding to a tree, we replace each undirected edge $(u, v)$ in the tree by the directed edges $(u, v)$ and $(v, u)$, so the resulting Eulerian tour of the graph is like a depth-first traversal of the tree without the need for of a distinguished root.  Henzinger and King~\cite{HK99} then showed how to implement such Euler tours for trees as dynamic balanced binary search trees, such that we can quickly support navigation queries and updates such as inserting an edge joining two trees and deleting an edge from a tree.  These implementations are called Euler-tour trees (ETTs).

\begin{figure*}
\begin{center}
\begin{tabular}{c@{\hspace{10ex}}c}
\includegraphics[width=.4\textwidth]{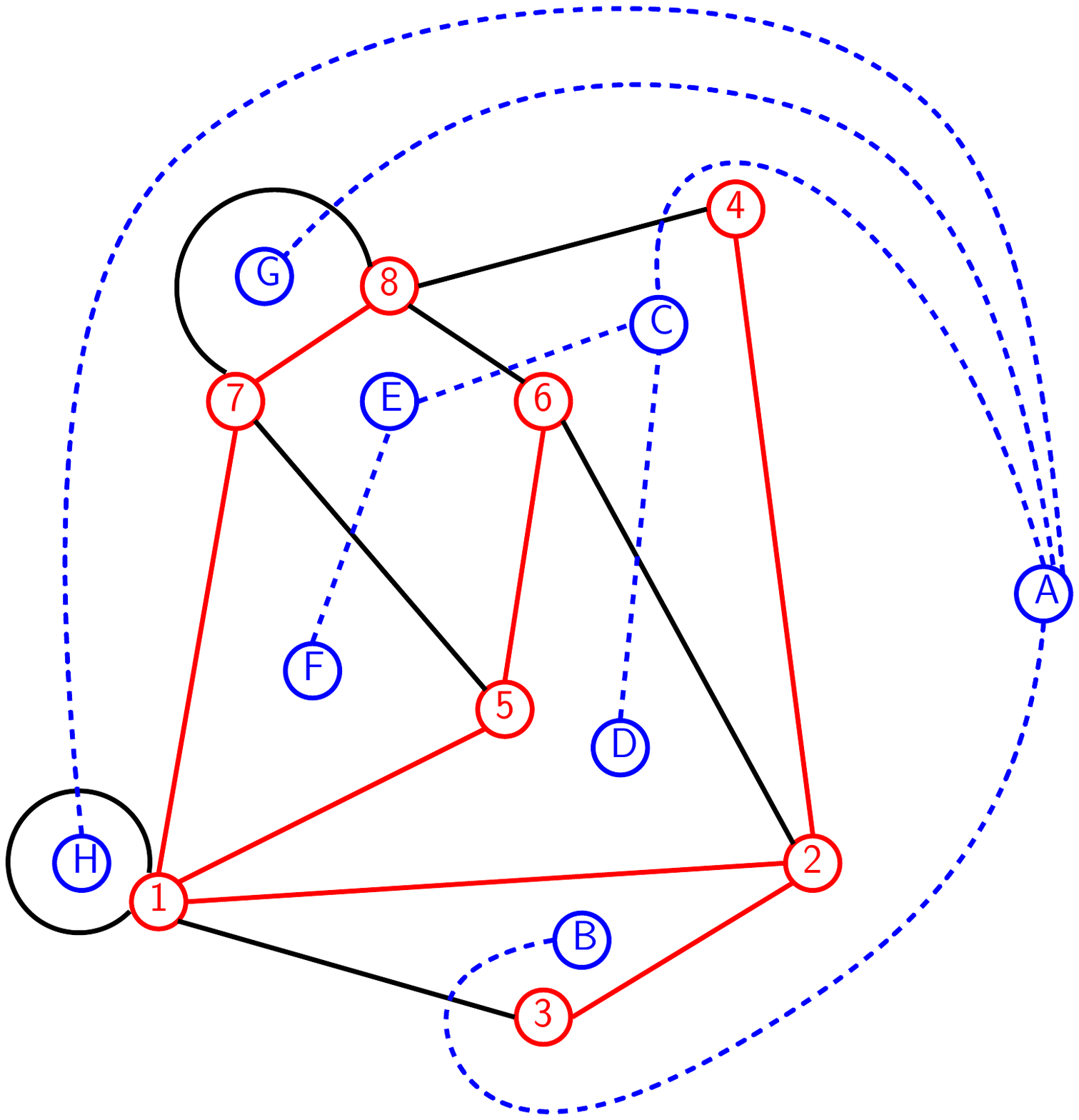} &
\includegraphics[width=.25\textwidth]{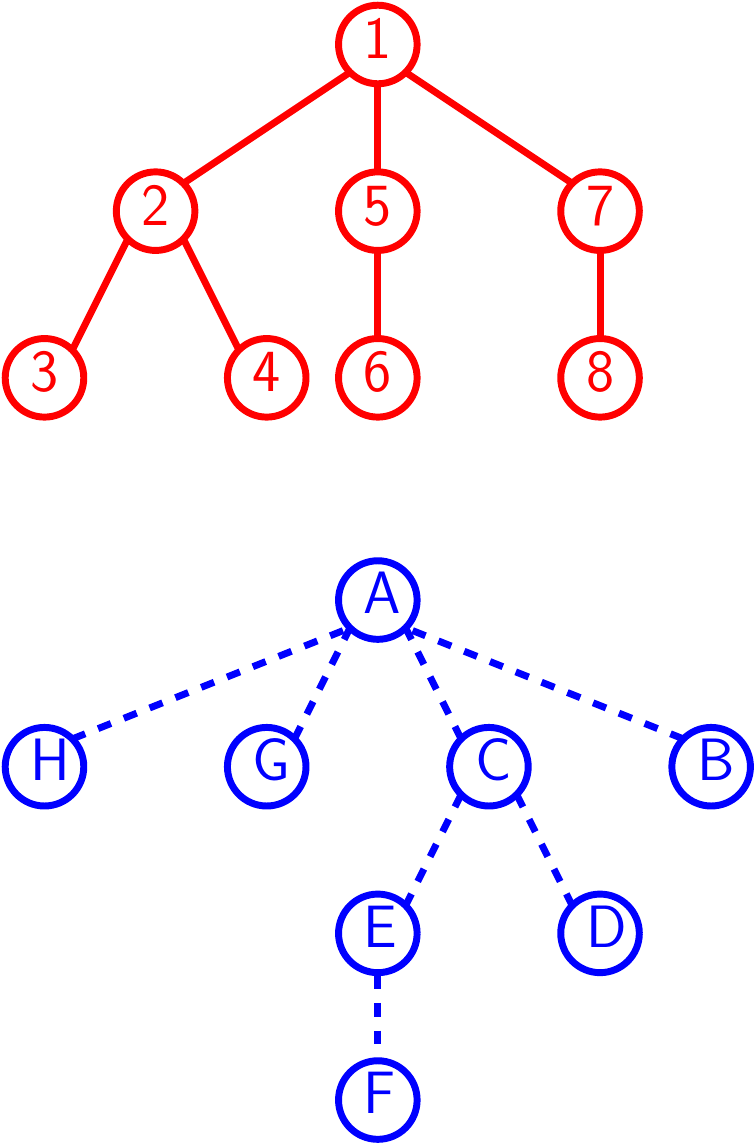}
\end{tabular}
\caption{Ferres et al.'s example of a planar graph {\bf (left)}, with the primal spanning tree {\bf (top right)} shown in red and the dual spanning tree {\bf (bottom right)} shown in blue.}
\label{fig:graph}
\end{center}
\end{figure*}

As far as we know, all current ETTs of a tree on $n$ vertices take $\Omega (n)$ words of space.  In contrast, we can store a {\em rooted} planar embedding of a tree (that is, an ordinal tree) on $n$ vertices in only $2 n + o (n)$ bits of space while still quickly supporting navigation queries and some updates, and such representations of ordinal trees are central to the field of compact data structures; see~\cite{Nav16} and references therein.  For example, Ferres et al.~\cite{FFGHN20} showed how to represent an embedding of a connected planar graph $G$ with $m$ edges in $4 m + o (m)$ bits by rooting and storing compactly a spanning tree of $G$ and an interdigitating spanning tree of the dual of $G$.  (The observation that any connected planar graph has such a pair of spanning trees appeared in Von Staudt's book {\it Geometrie de Lage}~\cite{Sta47} in 1847.)  Figure~\ref{fig:graph} shows the example from Ferres et al.'s paper, with the primal spanning tree shown in red and the dual spanning tree shown in blue.

\begin{figure*}
\begin{center}
\includegraphics[width=.6\textwidth]{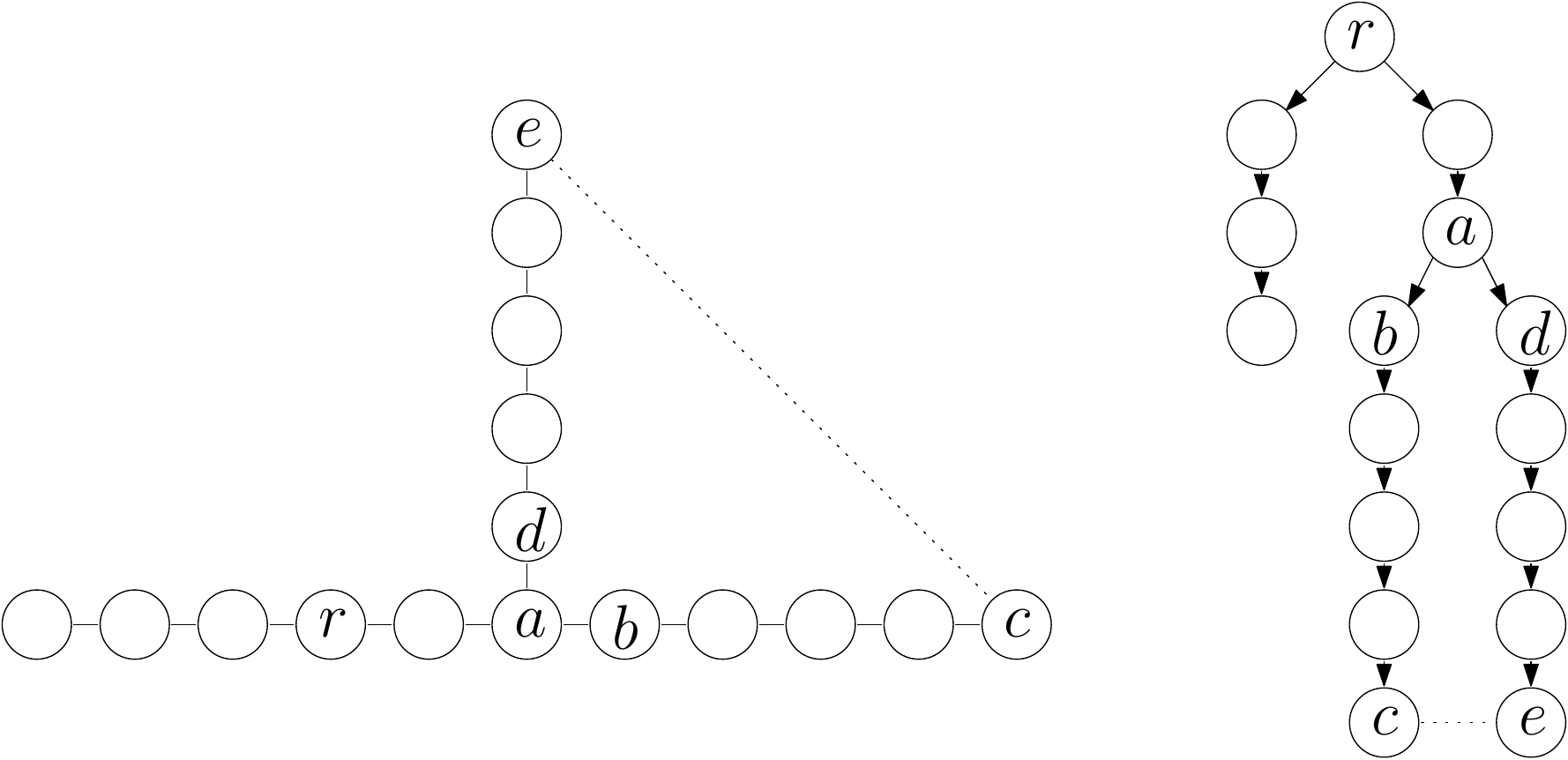}
\caption{A hard case for Ferres et al.'s data structure: however we choose to root the spanning tree of the graph {\bf (left)}, deleting one edge and inserting another~-- for example, if the root $r$ is in the left arm, then deleting $(a, b)$ and inserting $(c, e)$~-- forces us to reverse at least about $n / 3$ edges in the spanning tree {\bf (right)}.}
\label{fig:hard_case}
\end{center}
\end{figure*}

Ferres et al.'s data structure can be made to support quickly insertions of vertices and edges, but it does not adequately support deletions.  To see why, consider a graph on $n$ vertices with three arms of equal length, as shown on the left in Figure~\ref{fig:hard_case} (with the dotted edge $(c, e)$ not present initially).  The primal spanning tree must be the whole graph and, wherever we root it, two whole arms are branches.  Without loss of generality, suppose we root the spanning tree at $r$, so the paths from $b$ to $c$ and from $d$ to $e$ are branches in the spanning tree, as shown on the right.  If we delete the edge $(a, b)$ from the graph and insert the dotted edge $(c, e)$, then we have no choice but to reverse the directions in the spanning tree of all the edges either in the path from $b$ to $c$ or in the path from $r$ to $e$.  With the balanced-parentheses representation Ferres et al.\ used for their spanning trees, this takes $\Omega (n)$ time.  If we then delete $(c, e)$ and reinsert $(a, b)$, undoing the updates again takes $\Omega (n)$ time.

If the spanning tree in Figure~\ref{fig:hard_case} were represented by an ETT rather than with balanced paretheses, on the other hand, then deleting $(a, b)$ and inserting $(c, e)$ would be easy.  Indeed, we conjecture that it is possible to implement Ferres et al.'s data structure with ETTs such that all queries and updates take polylogarithmic time~-- but doing so would not currently be interesting because the entire data structure would no longer be compact.  Therefore, in this paper we start to investigate whether ETTs can be made compact while still quickly supporting a reasonable selection of queries and updates.  

In Section~\ref{sec:macro-trees} we describe a representation of ETTs that we call {\em macro-trees}, which slightly extend previous representations to allow weighted corners, where a {\em corner} is the ``gap between consecutive edges incident to some vertex''~\cite{HR17}.  For a forest $G$ on $n$ vertices, this representation takes $O (n)$ words of space and supports simple queries in constant time and more complex queries and updates in logarithmic time.

In Section~\ref{sec:micro-trees} we describe how if $G$ consists of a single tree with maximum degree $d \geq 2$ and $n$ is sufficiently large then, given a positive constant $\epsilon$, we can cluster the vertices such that each cluster contains between $\lg^{1 + \epsilon}(n)$ and $d \lg ^{1 + \epsilon}(n) + 1$ vertices.  We represent each cluster with a succinct {\em micro-tree}, such that all the clusters take a total of $2 n + o (n)$ bits, and then represent the $O (n / \log^{1 + \epsilon} n)$ inter-cluster edges with an $o (n)$-bit macro-tree.  The weight of each corner between two inter-cluster edges is the number of steps (all taken within the cluster) between those edges in the Euler tour of $G$.  

If $G$ consists of multiple trees, we cluster their vertices independently~-- with the bounds on the cluster sizes still depending on the maximum degree $d$ of the whole graph and the number $n$ of vertices it contains.  
This gives us our first result: we can store a collection of Euler-tour trees for a forest on $n$ vertices with maximum degree $d$ in $2 n + o (n)$ bits and support simple queries in constant time, more complex queries in $O (\log n)$ time and updates in $O (d \log^{1 + \epsilon} n)$ amortized time.
Since submitting this paper, we have realized how to remove the dependence on $d$ from the update time, as we describe in Section~\ref{sec:unbounded-degrees}.

Unless specified otherwise, all trees in this paper are taken to be planar embeddings of an unrooted tree (a.\,k.\,a.\ unrooted plane trees).
We assume throughout that $n$ is the total number of vertices in the maintained forest and we
are working in the word-RAM model with $\Theta (\log n)$-bit words.

\section{Macro-Trees}
\label{sec:macro-trees}

\begin{figure*}
\begin{center}
\includegraphics[width=.9\textwidth]{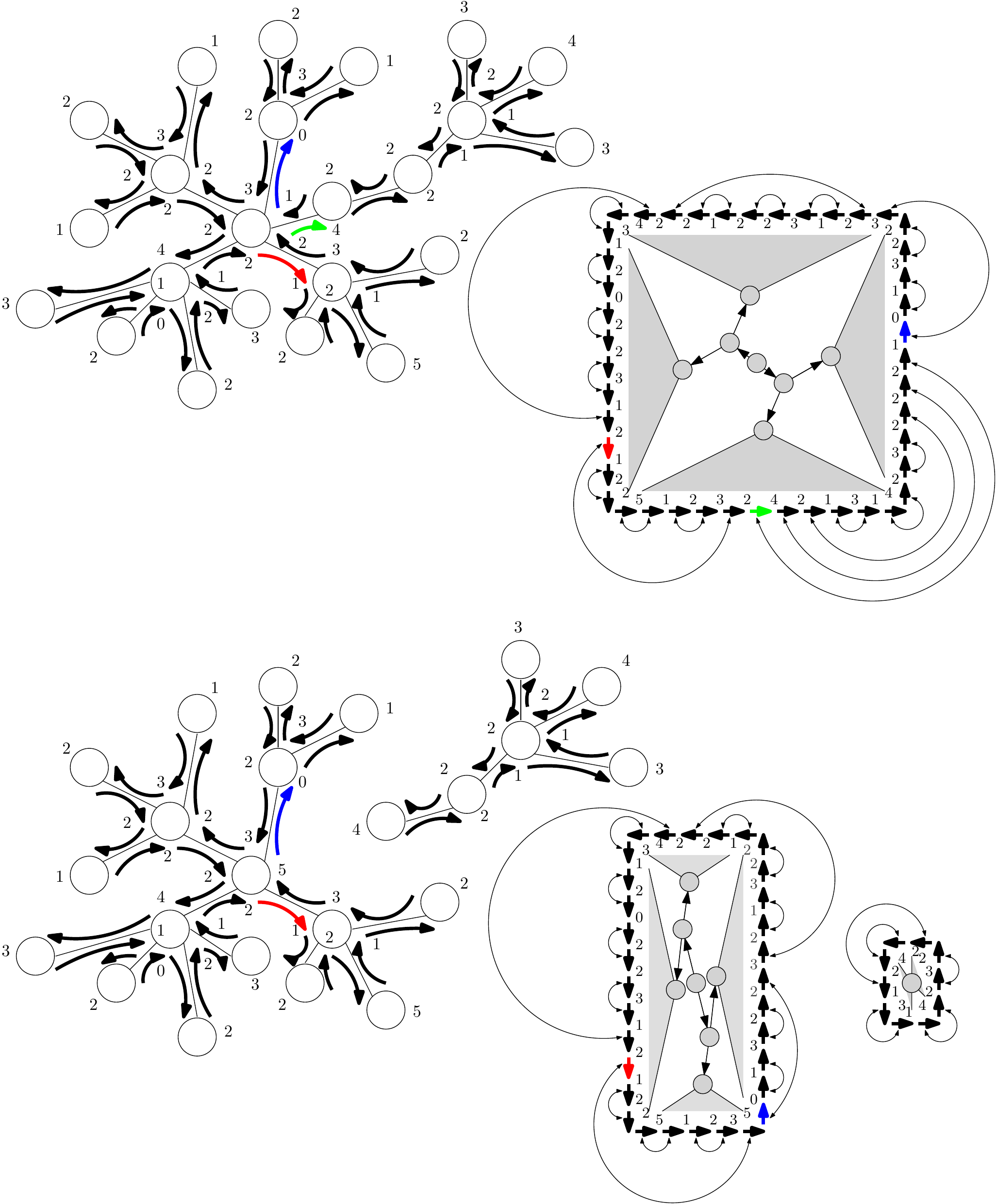}
\caption{An example of a macro-tree {\bf (top left)} and its representation {\bf (top right)}, and then the forest of two macro-trees obtained by deleting an edge {\bf (bottom left)}, and their representations {\bf (bottom right)}.}
\label{fig:macro-tree}
\end{center}
\end{figure*}

Suppose we are given a planar embedding of a forest $G$ with weighted corners, where each weight fits in a constant number of machine words; Figure~\ref{fig:macro-tree} shows an example at the top left, consisting of a single tree.  For each tree $T$ in $G$, we store $T$'s edges in a circular, doubly-linked list, in the order they are crossed in the Euler tour of $T$.  We also store a bidirectional pointer between each directed edge $(u, v)$ in $T$ and its reverse, $(v, u)$.

This allows us to move forward and backward in the Euler tour for $T$ one edge at a time, and to enumerate the edges incident to a given vertex $u$ in constant time per edge.  To see why, notice that if $(u, v)$ precedes $(u, w)$ in the counter-clockwise order of the edges incident to $u$, then $(v, u)$ precedes $(u, w)$ in the Euler tour.

Finally, we store an AVL tree whose leaves are the nodes in the list for $T$, and augment it such that given a directed edge $e$ and an integer $t$, in logarithmic time we can return the edge $e'$ such that the distance from $e$ to $e'$ in the list is $t$, or the total weight of the corners from $e$ and $e'$ is as close to $t$ as possible without being greater (or optionally less), or the sum of the distance and the total weight of the corners from $e$ to $e'$ is as close to $t$ as possible without being greater (or optionally less).

With the same AVL tree, in logarithmic time we can find the distance, or the total weight of the corners, in the list between two given edges of $T$.  This means, e.\,g., that we can quickly determine the size and total weight of the subtrees on either side of a given edge.

Figure~\ref{fig:macro-tree} includes an illustration of our representation at the top right, with the circular, doubly-linked list of directed edges shown as a square of arrows (with the pointer from each edge to its predecessor and successor in the list omitted for the sake of legibility); bidirectional pointers between directed edges $(u, v)$ and $(v, u)$ shown as arcs outside the square; and the AVL tree shown in grey inside the square, with the topmost nodes shown as circles with arrows from parents to children and then lower subtrees shown as triangles.

To advance from the red edge to the green edge, we follow the pointer to the reverse of the red edge, then move one position forward in the list; to advance from the green edge to the blue edge, we do the same thing.  To find the size of the subtree we traverse between the red edge and its reverse, we use the AVL tree to count the 6 directed edges between them in the list in logarithmic time, divide by 2 to get the number 3 of undirected edges in the tree, and add 1 to get the number 4 of vertices.  Similarly, we can sum the weights of the corners between the red edge and its reverse in the list.

We can change the weight of a corner in logarithmic time or, by splitting and joining doubly-linked lists and AVL trees, delete an edge in a tree represented by a macro-tree or insert an edge between vertices in two trees represented by macro-trees.  In our example, if we delete the undirected edge corresponding to the green edge and its reverse and assign the new corners weights 5 and 4, then we obtain two macro-trees shown at the bottom left of Figure~\ref{fig:macro-tree}, whose representations are shown at the bottom right.  Notice that now the blue edge follows the reverse of the red edge in the list.

\begin{lemma}
\label{lem:macro-trees}
Given a planar embedding of a forest $G$ on~$n$ vertices with weighted corners, we can store macro-trees for the trees in $G$ in a total of $O (n)$ words of space such that operations~(i) and (ii) take constant time and operations~(iii)\,--\,(xii) take $O (\log n)$ time:
\begin{enumerate}[label=(\roman*)]
\item given a directed edge $e$, return its predecessor and successor in the Euler tour of the tree containing $e$;
\item given a directed edge $(u, v)$, return its predecessor and successor in the counter-clockwise enumeration of edges incident to $u$;
\item given a directed edge $e$ and an integer $t$, return the directed edge $e'$ such that the distance from $e$ to $e'$ is $t$ in the Euler tour of the tree containing $e$;
\item given a directed edge $e$ and an integer $t$, return the edge $e'$ such that the total weight of the corners from $e$ to $e'$ in the Euler tour of the tree containing $e$ is as close to $t$ as possible without being greater (or optionally less);
\item given a directed edge $e$ and an integer $t$, return the edge $e'$ such that the sum of the distance and the total weight of the corners from $e$ to $e'$ in the Euler tour of the tree containing $e$ is as close to $t$ as possible without being greater (or optionally less);
\item given two directed edges $e$ and $e'$ in the same tree, return the distance and the total weight of the corners from $e$ to $e'$ in the Euler tour of that tree;
\item given an edge $e$, return the number of vertices and total weight of the corners in the subtrees on either side of $e$;
\item given a directed edge $e$ and a weight $w$, set to $w$ the weight of the corner after $e$ in the Euler tour of the tree containing $e$;
\item given an edge $e$ and weights $w$ and $w'$, delete $e$ from the tree containing it and set the new corners' weights to $w$ and $w'$;
\item given corners in two different trees $T$ and $T'$ and four weights, insert an edge between $T$ and $T'$ bisecting those corners and assign the four new corners the given weights;
\item given an edge $e$, contract $e$, thus fusing its endpoints and removing $e$ and its reverse edge from the Euler tour, adding up fused corner weights;
\item given two corners of the same vertex $v$, split $v$ into two nodes $v_1$ and $v_2$, connected by a new edge $e$, so that the neighborhoods of $v_1$ and $v_2$ result from the neighborhood of $v$ by splitting it at the given corners and inserting the new edge $e$ there.
\end{enumerate}
\end{lemma}

\section{Micro-Trees}
\label{sec:micro-trees}

For now, suppose $G$ consists of a single tree with maximum degree $d \geq 2$ and $n$ is sufficiently large.  Given a planar embedding of $G$ and a positive constant $\epsilon$, we can use essentially a centroid decomposition to partition $G$ recursively into clusters each containing between $\lg^{1 + \epsilon}(n)$ and $B = d \lg^{1 + \epsilon}(n) + 1$ vertices.

Suppose at some step of the recursion we are considering a subtree $S$ of $G$ on $n_S$ vertices such that $n_S > \lg^{1 + \epsilon}(n)$.  If $n_S \leq B = d \lg^{1 + \epsilon}(n) + 1$, then we can stop recursing.  Otherwise, we find a vertex or edge whose removal from $S$ leaves a forest in which each tree has size at most $n_S / 2$.

If we find such an edge $e$, then the two trees in the forest left by $e$'s removal from $S$ each have size
\[\frac{n_S}{2}
\ge \frac B2
> \lg^{1 + \epsilon} (n)\]
and so are large enough (and maybe too large) to be clusters; we recurse on them.  
If instead we find such a vertex $v$ then, since $v$ has degree at most $d$, at least one of the trees $S'$ in the forest left by $v$'s removal from $S$ has size
\[\frac{n_S - 1}{d}
> \lg^{1 + \epsilon}(n)\]
and so is large enough (and maybe too large) to be a cluster.  Since $S'$ contains at most $n_S / 2$ vertices, the rest of $S$ (including $v$) is a tree at least as big as $S'$, so it too is large enough (and maybe too large) to be a cluster.  We recurse on $S'$ and the rest of $S$.

Once we have partitioned $G$ into clusters, we consider that partition as a tree $P$ on $O (n / \log^{1 + \epsilon} n)$ vertices, with clusters in $G$ as vertices in $P$ and edges between clusters in $G$ as edges in $P$.  We store a macro-tree for~$P$, which takes $O (n / \log^{1 + \epsilon} n)$ words or $O (n / \log^{\epsilon} n) \subset o (n)$ bits, with the weight at a corner between $e$ and $e'$ in $P$ the number of steps between $e$ and $e'$ in the Euler tour of $G$.  We note that $P$ can have maximum degree more than $d$, but this does not affect Lemma~\ref{lem:macro-trees}.

\begin{figure*}
\begin{center}
\includegraphics[width=.5\textwidth]{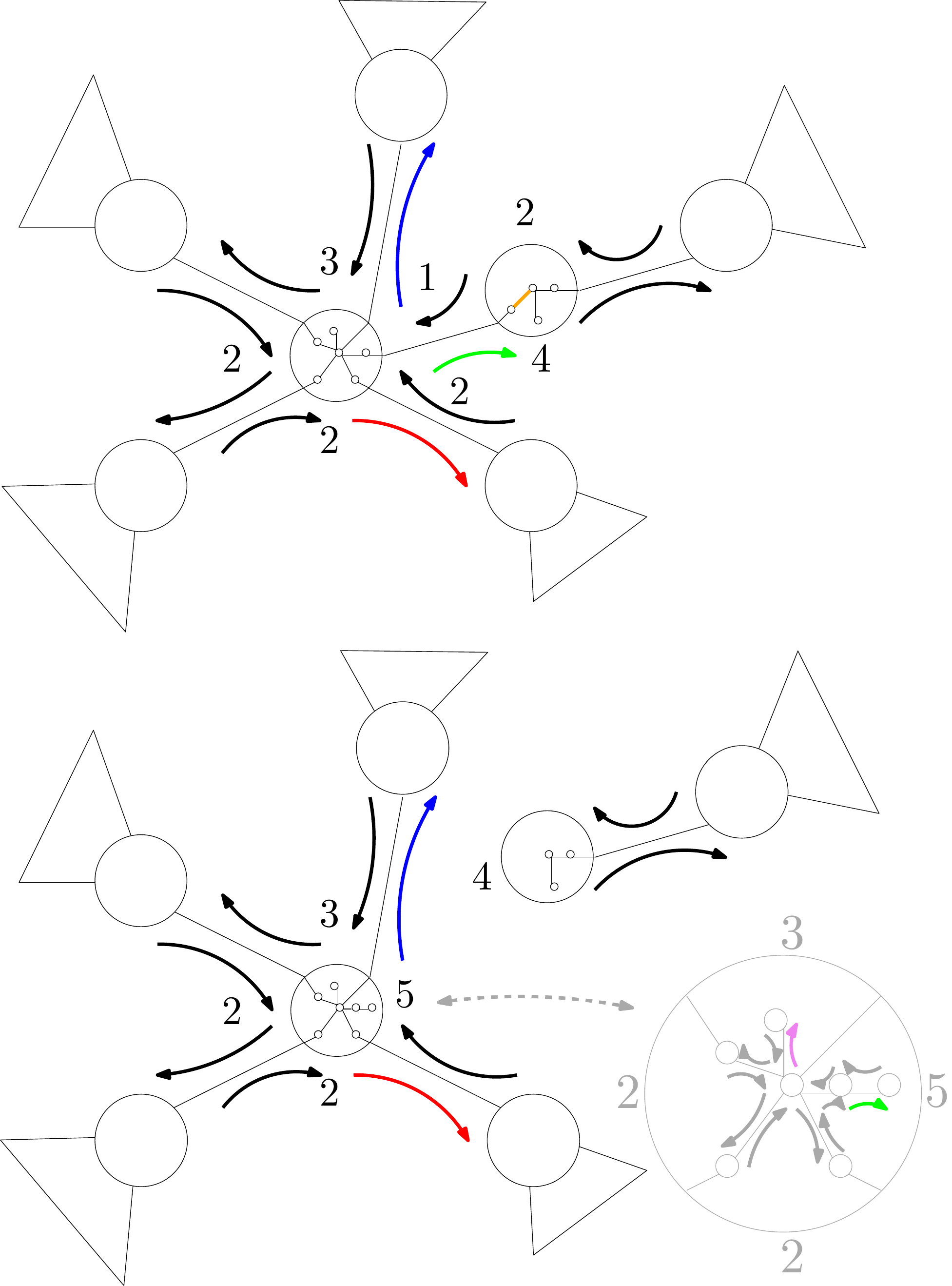}
\caption{An example of clusters, before the orange edge is deleted {\bf (above)} and after {\bf (below)}.  Deleting the orange undirected edge (adjacent to the green edge in the ``before'' example) means the green directed edge and its reverse retract into the other cluster (in the ``after'' example), as shown in the expanded view of that cluster {\bf (bottom right)}.}
\label{fig:clusters}
\end{center}
\end{figure*}

Figure~\ref{fig:clusters} shows an example of clusters embedded in vertices of the macro-tree from Figure~\ref{fig:macro-tree}, before and after an edge is deleted, with triangles representing subtrees.  The expanded view of one of the clusters shows why the corners incident to that cluster have weights 3, 2, 2 and 5: if we enter the cluster along the reverse of the blue edge, then we take 3 steps of the Euler tour inside the cluster before leaving the cluster again; when we re-enter for the first time, we take 2 steps inside before leaving again; when we re-enter for the second time, we also take 2 steps inside before leaving again; and finally, when we re-enter for the last time, we take 5 steps inside before leaving again, along the blue edge.  We note that the steps in the Euler tour to enter and leave the cluster do not count toward the weights of the corners.

We represent each cluster $C$ with a {\em micro-tree}: we temporarily ignore inter-cluster edges, root the remaining tree arbitrarily, and represent it succinctly as an ordinal tree using $2 n_C + o (n_C)$ bits, where $n_C$ is the number of vertices in $C$.  This takes a total of $2 n + o (n)$ bits over all the clusters.  Therefore, including the $o (n)$ bits for the macro-tree for $P$, thus far we are still representing $G$ using $2 n + o (n)$ bits.  However, we need to provide an interface between the inter-cluster edges and the micro-trees.

For a micro-tree on $n_C$ vertices, we store a ``ports bitvector'' with $2 n_C$ copies of \texttt{0} corresponding to the steps in a depth-first traversal of the micro-tree, with copies of \texttt{1} marking where inter-cluster edges are incident to vertices in the cluster (the cluster's ``ports'' to other clusters).  
The total length of these bitvectors is $2 n + O (n / \log^{1 + \epsilon} n)$ but only $O (n / \log^{1 + \epsilon} n)$ of the bits are \texttt{1}s, so compressed representations takes a total of $o (n)$ bits including support for rank and select~\cite{RamanRamanRao2007}.  We store a mapping from the inter-cluster edges ending in a cluster to the ranks of the \texttt{1}s in that cluster's ports bitvector, marking when in the depth-first traversal those inter-cluster edges touch vertices. We also store a mapping back from these \texttt{1}s to the inter-cluster edges in the macro-tree.  This takes $O (\log n \cdot n / \log^{1 + \epsilon} n) \subset o (n)$ bits, so we are still using $2 n + o (n)$ bits overall.

For example, if we root the cluster shown in the expanded view in Figure~\ref{fig:clusters} at the vertex reached by the green directed edge, on the right by the 5, then the bitvector is \texttt{0010001001001000}.  This means that if we start at our chosen root and walk counter-clockwise around the cluster, we take 2 steps before passing the first inter-cluster edge (pointing up and to the right), then 3 steps before passing our second (pointing up and left), then 2 before passing our third (pointing down and left), then 2 before passing our fourth (pointing down and right), and finally three more before reaching the root again.  Of course, if we view the bitvector as cyclic~-- corresponding to an Euler tour of the cluster rather than a depth-first traversal of the micro-tree~-- then the lengths of the runs of \texttt{0}s are the weights of the corners around that cluster in $P$.

If $G$ consists of multiple trees, we cluster them independently~-- with the bounds on the cluster sizes still depending on the maximum degree $d$ of the whole graph and the number $n$ of vertices in $G$.  If $G$ contains trees with less than $\lg(n)$ nodes, we store these all as a single dynamic string of balanced parentheses.  To be able to update the representations of the individual small trees, we keep the string divided into blocks of size roughly $\lg^{1 + \epsilon}(n)$ and completely replace any block we want to edit~-- much like the clusters.  The representation still takes $2 n + o (n)$ bits overall; since queries on those tiny trees trivially take $O (\log n)$ time, we obtain the same overall efficiency.

\subsection{Queries}

Suppose we know that the $i$th \texttt{0} in a cluster $u$'s bitvector indicates the step across a directed edge $e$ in the Euler tour of $u$, and the $j$th \texttt{0} in a cluster $v$'s bitvector indicates the step across a directed edge $e'$ in the Euler tour of $v$, and we want to compute the number of steps between $e$ and $e'$ in the Euler tour of $G$.  To do this, we first use rank and select queries on $u$'s and $v$'s bitvector to find the first \texttt{1} after the $i$th \texttt{0} in $u$'s bitvector and the last \texttt{1} before the $j$th \texttt{0} in $v$'s bitvector, the ranks of those \texttt{1}s and their distances from the $i$th and $j$th 0s.  This tells us how many steps in the Euler tour we take after crossing $e$ before we leave $u$ for the first time on an inter-cluster edge, and how many steps we take before crossing $e'$ after we enter $v$ for the last time on an inter-cluster edge.  We then map those \texttt{1}s to directed inter-cluster edges $(u, w)$ and $(x, v)$ and use Lemma~\ref{lem:macro-trees} to compute the distance and total weight of the corners between them in the Euler tour of the tree containing those edges.  The distance tells us the number of steps we take across inter-cluster edges between crossing $e$ and $e'$ in the Euler tour, and the total weight of the corners tells us the number of steps we take across intra-cluster edges between crossing $(u, w)$ and $(x, v)$.  Since we already know how many edges we cross between $e$ and $(u, w)$ and between $(x, v)$ and $e'$, we can compute the distance from $e$ to $e'$ in the Euler tour of the tree containing them.  This all takes $O (\log n)$ time, dominated by the time to query the AVL tree in the macro-tree representation to find out the distance and total weight of the corners between $(u, w)$ and $(x, v)$.

For example, suppose $e$ is the violet directed edge shown in the expanded view of the cluster in Figure~\ref{fig:clusters}, and $e'$ is the green edge in the same cluster (so $u = v$ in this case).  We cross 2 intra-cluster edges after $e$ before we leave the cluster across the inter-cluster edge pointing up and left, and cross 2 intra-cluster edges after re-entering the cluster across the inter-cluster edge arriving from down and right (the reverse of the red directed edge) before reaching $e'$.  The AVL tree at the bottom of Figure~\ref{fig:macro-tree} tells us there are 26 steps in the Euler tour of $P$ between when we leave the cluster heading up and left and when we re-enter it from down and right (including crossing those two inter-cluster edges), and the total weight of the corners between those directed inter-cluster edges is
\begin{eqnarray*}
&& 2 + 1 + 3 + 2 + 2 + 1 + 2 + 2 + 4 +\\
&& 3 + 1 + 2 + 0 + 2 + 2 + 3 + 1 + 2 +\\
&& 1 + 2 + 2 + 5 + 1 + 2 + 3\\
& = & 51\,,
\end{eqnarray*}
so the number of steps between $e$ and $e'$ in the Euler tour is
\[2 + 2 + 26 + 51 = 81\,,\]
not including crossing $e$ and $e'$ themselves.

Similarly, if we know the \texttt{0} in the bitvector of a cluster indicating a step across a directed edge $e$ and we are given an integer $t$, we can find the cluster containing the directed edge $e'$ that is $t$ steps after $e$ in the Euler tour of the tree containing $e$, and find the \texttt{0} in that cluster's bitvector indicting the step across $e'$, all in $O (\log n)$ time.  Using queries in the micro-trees and macro-tree we can also move forward or backward one step in any Euler tour in constant time, and enumerate the edges incident to a given vertex in constant time per edge.

Of course we cannot store identifiers of all the vertices in only $2 n + o (n)$ bits, so in general we need intra-cluster directed edges to be specified by which clusters they are in and the ranks of the \texttt{0}s indicating them, and vertices to be specified by specifying a directed edge leaving them.  We can afford to store mobile fingers to $O (n / \log^{1 + \epsilon} n)$ edges and vertices, however, without affecting our space bound.

\begin{lemma}
\label{lem:micro-trees}
Given a planar embedding of a forest $G$ on $n$ vertices, we can partition $G$ into clusters, store the partition as macro-trees and store each cluster as a micro-tree in a total of $2 n + o (n)$ bits, such that operations~(i) and (ii) take constant time and (iii)\,--\,(v) take $O (\log n)$ time:
\begin{enumerate}[label=(\roman*)]
\item given a directed edge $e$, return its predecessor and successor in the Euler tour of the tree containing $e$;
\item given a directed edge $(u, v)$, return its predecessor and successor in the counter-clockwise enumeration of edges incident to $u$;
\item given a directed edge $e$ and an integer $t$, return the directed edge $e'$ such that the distance from $e$ to $e'$ is $t$ in the Euler tour of the tree containing $e$;
\item given two directed edges $e$ and $e'$, return the distance from $e$ to $e'$ in the Euler tour of the tree containing them;
\item given an edge $e$, return the number of vertices in the subtrees on either side of $e$.
\end{enumerate}
\end{lemma}

Notice that we do not mention $G$'s maximum degree $d$ in Lemma~\ref{lem:micro-trees}.  This is because $d$ appears only in the upper bound $B$ on clusters' sizes, which does not affect query times but only updates, which we discuss next.

\subsection{Updates}
\label{sec:updates}

Recall that $B=d \lg^{1 + \epsilon} (n)+1$ here.
To insert an edge between two trees or delete an edge in a tree, we completely rebuild the micro-trees, bitvectors and mappings for the affected clusters, in $O (B)$ time~-- possibly choosing new roots for the new micro-trees~-- and update the macro-tree or macro-trees in $O (\log n)$ time.  
We may need to split or join a constant number of clusters to maintain the invariant that they all have between $\lg^{1 + \epsilon} (n)$ and $B$ vertices, but this can still be handled in $O (B)$ time using standard techniques.  A more drastic problem occurs when so many vertices are added or deleted that the bounds for our cluster sizes change and we must rebuild many clusters, but this cost can be amortized over the insertions and deletions.

To delete the orange edge in Figure~\ref{fig:clusters}, we move the shared endpoint of the orange and green edges into the same cluster as the other endpoint of the green edge, then completely rebuild the micro-trees, bitvectors and mappings for the clusters shown inside the vertices of the macro-tree.  We update the macro-tree by deleting the green edge and its reverse from the macro-tree (since they are now intra-cluster edges), and weighting the new corners by the number of steps in the Euler tours of the clusters between the preceding and succeeding edges: the cluster that contained the orange edge now has 3 vertices, 2 (undirected) intra-cluster edges and 1 (undirected) inter-cluster edge, so the number of steps in the Euler tour of $T$ between entering it and leaving it is 4, so that is the weight of its new single corner; the cluster that now contains the green edge now has 7 vertices, 6 (undirected) intra-cluster undirected edges and 4 (undirected) inter-cluster edges, and the number of steps in the Euler tour between entering it across the last inter-cluster edge before the green edge and leaving it across the first inter-cluster edge after the green edge is 5, so that is the weight of its new corner.

\begin{lemma}
\label{lem:updates}
After deleting a given edge from a tree in a forest on a total of $n$ vertices with maximum degree $d$, we can update our representation of that tree in $O (B)$ amortized time and obtain the representations of the two resulting trees.  Similarly, after inserting an edge bisecting given corners in two trees of a forest on a total of $n$ vertices with maximum degree~$d$, we can update our representations of those trees in $O (B)$ amortized time and obtain a representation of the single resulting tree.  We can add or delete an isolated vertex in $O (B)$ amortized time.
\end{lemma}

Combining Lemmas~\ref{lem:micro-trees} and~\ref{lem:updates}, we obtain our first theorem:

\begin{theorem}
\label{thm:main}
Given a planar embedding of a forest $G$ on $n$ vertices with maximum degree $d$, we can store $G$ in $2 n + o (n)$ bits such that operations (i) and (ii) take constant time, operations~(iii)\,--\,(v) take $O (\log n)$ time and (vi) and (vii) take $O (d \log^{1 + \epsilon}(n))$ time:
\begin{enumerate}[label=(\roman*)]
\item given a directed edge $e$, return its predecessor and successor in the Euler tour of the tree containing $e$;
\item given a directed edge $(u, v)$, return its predecessor and successor in the counter-clockwise enumeration of edges incident to $u$;
\item given a directed edge $e$ and an integer $t$, return the directed edge $e'$ such that the distance from $e$ to $e'$ is $t$ in the Euler tour of the tree containing $e$;
\item given two directed edges $e$ and $e'$ in the same tree, return the distance from $e$ to $e'$ in the Euler tour of that tree;
\item given an edge $e$, return the number of vertices in the subtrees on either side of $e$;
\item given an edge $e$, delete $e$ from the tree containing it and return the representations of the two resulting trees;
\item given corners in two different trees $T$ and $T'$, insert an edge between $T$ and $T'$ bisecting those corners and return the representation of the resulting tree.
\end{enumerate}
\end{theorem}

\section{Trees of Arbitrary Degree}
\label{sec:unbounded-degrees}

The above scheme for bounded-degree forests can be generalized to arbitrary forests
by splitting high-degree vertices.
In the following, we describe the necessary changes to the data structure.

\subsection{Macro-Trees}

We modify the macro-tree by allowing (inter-cluster) edges to be either ``true'' or ``false''.
A true edge is as before, whereas a false edge does not actually correspond to any edge in $G$,
but rather connects two clones of the same graph vertex in different clusters.
In the implementation, we can identify each edge and its preceding corner into a single entity with a weight, a ``macro-edge'';
a false edge adds weight 0 and a true edge adds weight 1 to the macro-edge. 
Conceptually, edges in the macro-tree are
(potentially empty) sequences of (consecutive) true edges in the Euler tour, 
plus optionally one false edge at the end of such a sequence.

As before, macro-edges are kept in a linked list, with pointers to their reverse traversals.
We also add pointers to the immediate true successors and predecessors of each macro-edge.
Any balanced BST that supports splitting and merging, augmented with subtree weights,
can be used to implement efficient access to macro-edges.
Operations stay the same, except that false edges have to be counted with weight~0.

\subsection{Micro-Trees}

Our decomposition for unbounded degrees follows a similar approach as above, but caters for large degrees by splitting vertices.
Given a total size $n$ and a constant $\epsilon>0$, we now set $B = 3\lg^{1+\epsilon}(n)$.
We decompose a tree $T$ on $n_T$ vertices as follows:
If $n_T \le B$, it forms a cluster of its own.
Else, we find a centroid, i.e., a node or edge, so that after its removal, all remaining subtrees have size at most $\frac12 n_T$. 
If we find a centroid edge in $T$, we recurse as before.
Otherwise we find a node $v$ that splits $T$ into subtrees $S_1,\ldots,S_d$, for $d=\deg(v)$,
of sizes $n_{S_1},\ldots,n_{S_d} \le \frac12 n_T$.
If any of these subtrees has size $n_{S_i} \ge B / 3$, 
we recursively decompose $S_i$ and the rest of the tree.
Otherwise, if $n_{S_1},\ldots,n_{S_d} < \frac13 B$, let $j$ be the index that minimizes 
\[\left| (n_{S_1}+\cdots+n_{S_j}) - (n_{S_{j+1}}+\cdots+n_{S_d})\right| < B / 3.\]
We split $v$ into two ``clones'', $v_1$ and $v_2$, and give
$v_1$ the neighbors $S_1,\ldots,S_j,v_2$ and
$v_2$ the neighbors $v_1,S_{j+1},\ldots,S_d$.
The edge $\{v_1,v_2\}$ is marked as a ``false'' edge
and separates $T$ into two components, which are recursively decomposed.

If we start with a tree $T$ with $n_T \ge B$, 
then all clusters have between $\frac13 B=\lg^{1+\epsilon}(n)$ and $B$ nodes,
and the number of clusters is $\Theta(n/B)$.
Note that intra-cluster edges are always true edges, hence
the representation of clusters using micro-trees remains unaffected.

We thus obtain the same result as in Theorem~\ref{thm:main}
for general forests:

\begin{theorem}
\label{thm:main-unbounded}
Given a planar embedding of a forest $G$ on $n$ vertices, we can store $G$ in $2 n + o (n)$ bits such that operations (i) and (ii) from Theorem~\ref{thm:main} take constant time, operations~(iii)\,--\,(v) take $O (\log n)$ time and (vi) and (vii) take $O (\log^{1 + \epsilon}(n))$ time.
\end{theorem}

\section{Acknowledgments}

The first author was funded by NSERC Discovery Grant RGPIN-07185-2020.

\end{document}